\documentclass[preprint,12pt]{elsarticle}

\usepackage{amssymb}
\journal{}

\begin{document}

\begin{frontmatter}



\title{Bayesian Nonparametric Survival Analysis using mixture of Burr XII distributions}


\author{S. B. Hajjar}

\address{bohlurihajjar.soghra@razi.ac.ir}
%\email{bohlurihajjar.soghra@razi.ac.ir}
\author{ S. Khazaei}
\address{s.khazaei@razi.ac.ir}

\begin{abstract}
Recently, the Bayesian nonparametric approach in survival studies attracts much more attentions. Because of multi modality in survival data, the mixture models are very common in this field. One of the famous priors on Bayesian nonparametric models is Dirichlet process prior. In this paper we introduce a Bayesian nonparametric mixture model with Burr distribution(Burr type XII) as the kernel of mixture model. Since the Burr distribution shares good properties of common distributions on survival analysis, it has more flexibility than other distributions. By applying this model to simulated and real failure time data sets, we show the preference of this model and compare it with other Dirichlet process mixture models with different kernels. And also we show that this model can be applied for the right censored data. For calculating the posterior of the parameters for inference and modeling, we used the MCMC simulation methods, especially Gibbs sampling.
\end{abstract}


\end{frontmatter}


\section{Introduction}
Many distributions are commonly used for modeling failure time data, whether for reliability studies in manufacturing or survival analysis in the health domain. These include the exponential, Weibull, log-normal and log-logistic distributions. The log-normal distribution is an attractive distribution for modeling component failure times because it has non-monotone failure rates. Also, the log-logistic distribution has been used because it shares this property. The Burr(XII) distribution has recently emerged as a promising distribution for use with failure time data (\cite{rao},\cite{lanjoni}). It shares many properties with the more traditional distributions used with failure time data: the log-logistic distribution is a special case of the Burr(XII) distribution; the Burr(XII) distribution can be a good approximation to the Weibull distribution; and the Weibull distribution is the limiting distribution of the Burr(XII) distribution \cite{tadikamalla}. While the Burr(XII) distribution shares many of the advantages of these other distributions, it is more flexible.

While parametric distributions offer convenience in modeling, it is often difficult to choose and/or justify an appropriate parametric form for lifetime data. This may be particularly difficult with complex systems with multimodal survival times or when modeling failure time data from products manufactured using emerging technologies with unfamiliar mechanisms \cite{cheng13}. Such multimodal data can be more flexibly modeled using a mixture model rather than a single parametric distribution. The mixture model can be written as a summation of a coefficient and the probability of choosing that coefficient (kernel) \cite{Mclachlan}. The Bayesian mixture model is obtained by allowing these coefficients to be random. The nonparametric Bayesian mixture model gains even greater flexibility by considering the space of parameters to be infinite.

Nonparametric Bayesian mixture models can be characterized by the choice of mixture distribution. One of the most common choices for the distribution of coefficients is the Dirichlet process, which yields the Dirichlet Process mixture model(DPMM) \cite{cheng13}. The DPMM can be further tailored to the application through the choice of distributional form for the kernel. A DPMM with a normal kernel is often used (\cite{escobar95}, \cite{Muller}); however, it is unsuitable for survival data given the positive support required for failure times. A DPMM with a Weibull kernel has been studied for survival analysis by Kottas \cite{kottas6} and a DPMM with a log-normal kernel was developed for reliability applications by Cheng and Yuan \cite{cheng13}. This work proposes the use of a DPMM with a Burr(XII) distribution for the kernel which also accommodates the positive support required for failure time data.

In the next section we describe the general framework for DPMMs. Section 3 describes the use of the Burr(XII) kernel in the DPMMs. The Gibbs sampling estimation algorithm for the DPMM with Burr(XII) kernel will be described in Section 4. Section 5 presents simulation results and an application of the DPMM with Burr(XII) kernel to failure time data. Finally, the obtained results are discussed in section 6.

\section{Dirichlet process mixture models}

\subsection{Dirichlet Process}
Let $(\Theta,\mathcal{A},G)$ be a probability space, let $G_0$ be a distribution over $\Theta$ and $\upsilon$ be a positive real number, then for any finite measurable partition $A_1,A_2,...,A_k$ of $\Theta$, G will be called a Dirichlet process (DP) with base distribution $G_0$ and concentration parameter $\upsilon$, $G\sim DP(\upsilon,G_0)$, if for every vector $\bigl(G(A_1),G(A_2),...,G(A_k)\bigr), k\in N$, we have
$$\Bigl(G(A_1),G(A_2),...,G(A_k)\Bigr)\sim Dir(\upsilon G_0(A_1),...,\upsilon G_0(A_k)).$$

An important constructive definition of a DP is given by Sethuraman \cite{sethuraman94} and it is based on the discrete nature of the
process. In this construction, $\theta_h$'s are i.i.d from the centering measure $G_0$ and each weight $w_h$ is defined as a fraction of $1-\sum_{l<h}w_h$.
Let $w_h=\nu_h\prod_{l<h}(1-\nu_l)$ where $\nu_h\begin{array}{c}
     \tiny{i.i.d} \vspace{-0.3cm} \\
     \sim \\
     \end{array}Be(1,\upsilon)$, $\theta_h\begin{array}{c}
     \tiny{i.i.d} \vspace{-0.3cm} \\
     \sim \\
     \end{array}G_0$, and also assume $\upsilon$ and $\theta_h$ are independent, then
$$G(.)=\sum_{h=1}^{\infty} w_h\delta_{\theta_h}(.),$$
define a DP($\upsilon,G_0$) random probability measure. This representation of a DP is known as "Stick-Breaking" representation.

\subsection{Dirichlet Process Mixture Models}
A parametric mixture model for
function $f(x)$ can be written as
\begin{equation}
f(x|\pi_1,...,\pi_M,\theta_1,...,\theta_M)=\sum_{j=1}^{M}k(x|\theta_j)\pi_j;~\theta_j \in \Theta,~j=1,...,M,
\end{equation}
where $k(x|\theta)$ is a parametric kernel and $\pi_j$'s are the values that for all
j; $\pi_j>0$ and $\sum_{j=1}^{M}\pi_j=1$.

By the Bayesian approach
%by the Bayesian approach, the weights ($\pi_j$'s) as the probability of choosing $\theta_i$'s can be considered, and
we can consider the weights($\pi_j$'s) as the probability of choosing $\theta_j$'s and then by putting a nonparametric Bayesian prior like Dirichlet process on the wights, we have a mixture model named by Dirichlet process mixture models(DPMMs). A DPMM can be illustrated as bellow:
$$f_G(x)=\int_{\Theta}k(x|\theta)dG(\theta)$$
where $\theta\sim G$ and $G\sim DP(\upsilon,G_0)$.

By a hierarchical representation, a DPMM can be presented as the following form, which is more popular of these type models,
\begin{eqnarray}
x_i|\theta_i&\sim & k(x_i|\theta_i)\quad i=1,...,n \nonumber\\
\theta_i|G&\sim & G(\theta)\\
G&\sim & DP(\upsilon,G_0).\nonumber
\end{eqnarray}
Through this hierarchical representation, a marginalization step is often used in implementation of the model. Blackwell and McQueen \cite{blackwell} showed that by integrating out of G, the joint distribution of $\theta_1,...,\theta_n$ may be written into a product of conditional distribution of the following form
\begin{equation}
\theta_i|\theta_1,...,\theta_{i-1}, G_{0,\upsilon}\sim \frac{1}{i-1+\upsilon}\sum_{j=1}^{i-1}\delta_ {\theta_j} + \frac{\upsilon}{i-1+\upsilon}G_0,~~~~i=1,...,n
\end{equation}
where $\delta(\theta)$ denotes the distribution concentrated at point $\theta$, and as we see, this formula has the form  a mixture model.
\subsection{Model implementation}
Our aim here is to formulate how to sample from the DPMMs by Gibbs sampling. According to \cite{kottas6}, Gibbs sampling for drawing sample from $[(\theta_1,...,\theta_n),\upsilon,...|t]$ is based on the following full conditionals (we use the bracket([]) to show the conditional and marginal distributions):
\begin{eqnarray}
&& (1)~ [(\theta_i)|(\theta_{-i},z_{-i}),\upsilon,...,t],\quad for ~i=1,...,n\nonumber\\
&& (2)~ [(\theta_j^*)|z,n^*,\upsilon,...,t], ~~for~ j=1,...,n^*\\
&& (3)~ [\upsilon|\{(\theta_j^*),j=1,...,n^*\},n^*,t],[...|\{(\theta_j^*),j=1,...,n^*\},n^*,t].\nonumber
\end{eqnarray}
where t is the failure time data.
Here, $\theta_i$'s are the parameters of kernel in DPMMs that we want to analyze them. For example if the kernel of a DPMM is two-parameter Burr(XII) distribution with (c,k) parameters, as it will be discussed by details in the next section, then $\theta=(\theta_1,\theta_2)=(c,k)$.
The model (3) and discreteness property of DP, exhibit a clustering kind of effect.
%we have the clustering in $\theta_i$'s.
\newline We present $n^*$ as the number of the clusters between $\theta$'s that denote by $\theta^*$'s.
The vector of indicators $z=(z_1,...,z_n)$ indicates the clustering configuration such that, $z_i=j$ when $\theta_i=\theta_j^*$.
Also, the $\theta_{-i}$ that used in (4), it will be defined by $\theta_{-i}=(\theta_1,\theta_2,...,\theta_{i-1},\theta_{i+1},...,\theta_d)$.

By combining formula (3) with the likelihood of data\cite{Muller}, the posterior distribution can be presented as
\begin{equation}
f(\theta_i|\theta_{-i},t_i)=b\upsilon f(t_i;\theta_i)G_0(\theta)+b\sum_{j=1,j\neq i}^nf(t_i;\theta_i)\delta_{\theta_j^*}(\theta_j)
\end{equation}
where
$$b=(q_0+\sum_{j=1,j\neq i}^n f(t_i;\theta_i))^{-1}$$
and
$$q_0=\upsilon\int f(t_i|\theta)G_0(\theta)d\theta.$$
Note that by \cite{cheng13}, we have the posterior of $\theta_i$ as
$$h(\theta_i|t_i)=\frac{f(t_i;\theta_i)G_0(\theta_i)}{\int f(t_i|\theta)G_0(\theta)d\theta},$$
then we can rewrite equation (5) as follows
\begin{equation}
f(\theta_i|\theta_{-i},t_i)=bq_0 h(\theta_i|t_i)+b\sum_{j=1,j\neq i}^nf(t_i;\theta_i)\delta_{\theta_j^*}(\theta_j).
\end{equation}
Regarding the form of the model (5) or (6), we can draw from the distribution by Gibbs sampling such that $f(\theta_i|\theta_{-i},t_i)$ takes the one of the previous values with probability of $bf(t_i;\theta_i)$ and takes a new $\theta$ from $h(\theta_i|t_i)$ with probability of $bq_0$.

\subsection{Simulation-based parameter estimation algorithm}
The ability of MCMC methods to compute integrals caused that statisticians widely use it in the implementation of the Bayesian approach.
An MCMC method constructs a Markov chain which is a stochastic process that the current value generated from step will always depend on the value in the previous step.
This algorithm is a preferable method to sample from conditional distributions in Bayesian approach.
%Sampling with this method can be explain as following algorithm
We may summarize this method as follows
\begin{eqnarray*}
&&\hspace{-7cm}Algorithm: Gibbs~sampler\\
&&\hspace{-6cm} 1.~ Initialize~~\theta^{(0)}\sim f(\theta)\\
&&\hspace{-6cm} 2.~ For~iteration~~i=1,2,...~do\\
&&\hspace{-6cm} ~~~~~~\theta_1^{(i)}  \sim f(\theta_1|\theta_2^{(i-1)},\theta_3^{(i-1)},...,\theta_d^{(i-1)},D),\\
&&\hspace{-6cm} ~~~~~~\vdots  \\
&&\hspace{-6cm} ~~~~~~ \theta_d^{(i)} \sim f(\theta_d|\theta_1^{(i)},\theta_2^{(i)},...,\theta_{d-1}^{(i)},D),
\end{eqnarray*}
where $\theta_1,...,\theta_d$ represent the parameters in the model and D is the observation set. The values of iteration i would be sampled from the last version of the other values.

In our model with Dirichlet process prior for the infinite parameters, Gibbs sampling is more appropriate simulation method.

\section{Dirichlet process Burr(XII) mixture model}
The Burr distribution is the renowned distribution in the probability. Burr\cite{burr42} has propose a numbers of forms of c introduced a system of distributions by considering distribution functions F(x) satisfying the differential equation
$$\frac{dF}{dx}=A(F)g(x).$$
In the special case if
 $A(F)=F(1-F)$ then the solution of Burr's equation is
$$F(x)=\frac{1}{1+exp(-G(x))}$$
where $G(x)=\int_{-\infty}^{x}g(t)dt.$
Twelve distributions that named by Burr distributions are particular form of this solution.
Here we use the type XII of Burr distributions that is named by Burr(XII) distribution \cite{rodriguez}.

Let $k_B(t|c,k)$ and $K_B(t|c,k)$ denote the p.d.f and c.d.f
of the Burr(XII) distribution respectively, which defined by
$$k_B(t|c,k)=ck\frac{t^{c-1}}{(1+t^c)^{k+1}}\qquad c,k>0,~ t>0$$
and
$$K_B(t|c,k)=1-(1+t^c)^{-k}$$
which the parameter space is $\Theta=\{(c,k); 0<c<\infty , 0<k<\infty\}.$
Suppose base distribution $G_0$ as the prior guess for the joint distribution of $c$ and $k$.
 When we choose with Burr(XII) kernel, the $G_0$ that yields closed form expression for $\int k_B(.|c,k) G_0(dc,dk)$ is not available, but we choose a multiple distributions of uniform$(0,\phi)$ and exponential with parameter $\gamma$ for the $G_0$,
\begin{equation}
G_0(c,k|\phi,\gamma)=Unif(c|0,\phi)\times Exp(k|\gamma).
\end{equation}
This choice achieves aforementioned goals. Suppose $\gamma$ and $\phi$ are random and have $Pareto(a_\phi,b_\phi)$ and $IGamma(a_\gamma,b_\gamma)$
distributions respectively (Here $IGamma(.|a,b)$ denotes the inverse-gamma distribution with mean $\frac{b}{a-1}$ provided $a >1).$
\newline We set the $a_\phi=a_\gamma=d$ and then we choose d=2, since this value makes the variance of Pareto distribution infinite that cover all values in R. We determine the values of $b_\phi$ and $b_\gamma$ by the data that the details are presented in the appendix A.

Finally, by considering Burr(XII) as the kernel of the DPMM we introduce the hierarchical model of Dirichlet process Burr(XII) mixture model (DPBMM) as the following form
\begin{eqnarray}
t_i|c_i,k_i&\sim & Burr(t_i|c_i, k_i),\quad i=1,...,n,\nonumber\\
(c_i,k_i)|G &\sim & G,\nonumber\\
G &\sim & DP(\nu,G_0),\\
\nu, \gamma, \phi &\sim& Gamma(a_\nu,b_\nu)\times IGamma(a_\gamma,b_\gamma)\times Pareto(a_\phi,b_\phi).\nonumber\\
\nonumber
\end{eqnarray}

\section{Modeling based on DPMMs with Burr(XII) kernel}
Now we describe how to estimate parameters with the DPBMMs.
Data contains rightly censored failure times that are indicated by a characteristic function:
$$\delta_i=\{^{\textrm{0\qquad if $t_i$ is an  uncensored failure time}}_{\textrm{1\qquad if $t_i$ is a censored failure time}}.$$
As is denoted before, $n^*$ is the number of distinct cluster of parameters $(c_i,k_i)$ in which $i=1,...,n$ and let $(c^*_j,k^*_j), ~j=1,...,n^*$ denote the cluster locations.
Then the clustering indicator $z_i$ indicates the cluster which $(c_i,k_i)$ belongs to
$z_i=j$ if $(c_i,k_i)=(c^*_j,k^*_j)$. Let $n^*_j$ be the number of members of cluster j.

A simulation algorithm to implement Gibbs sampling is drawing a sample for one parameter from its conditional posterior density conditioned
on the last estimates of all the other parameters[12].
Kottas \cite{kottas6} is derived the conditional posterior densities for the Dirichlet process Weibull mixture model. So by replacing the Weibull with Burr(XII) as the kernel, we derive the conditional posterior densities in DPBMMs.

At first we draw a sample of $(c_i,k_i)$ and update the $z_i$ for each failure time $t_i$. The sample can be a new value from $G_0$ or be equal to an existing values.
Let $n^{*(i)}$ be the number of clusters when $(c_i,k_i)$ is removed from the sample and let $n^{*(i)}_j$ be the number of elements in cluster j after removing.

We denote the quantities related to an observed data by letter "o" and censored data with letter "c".
If $t_{io}$ is an observed (uncensored) failure time, then by \cite{kottas6} and equation (6) the conditional posterior density of $(c_i,k_i)$ is the mixed distribution
$$f(c_i,k_i|{(c_i,k_i);i\neq i'},\nu, \gamma, \phi,t_{io})=\frac{q^o_0 h^o(c_i,k_i|\phi,\gamma,t_{io})+\sum_{j=1}^{n^{*(i)}}n^{*(i)}_j q^o_j\delta_{c^*_j,k^*_j}}{q_0^o+\sum_{j=1}^{n^{*(i)}}n^{*(i)}_j q^o_j},$$
where $q_j^o=k_B(t_{io}|c_j^*,k_j^*)$ and $q_0^o$ in is given by
\begin{eqnarray*}
q_0^o &=& \nu \int_0^\phi \int_0^\infty k(t_{io}|c,k)G_0(c,k)dcdk\\
&=& \frac{\nu}{\phi} \int_0^\phi \frac{ct_{io}^{c-1}}{(1+t_{io}^c)}(\int_0^\infty\frac{ke^{-\frac{k}{\gamma}}}{(1+t_{io}^c)^k}dk)dc\\
&=& \frac{\nu}{\phi} \int_0^\phi \frac{ct_{io}^{c-1}}{(1+t_{io}^c)(ln(1+t_{io}^c)+\frac{1}{\gamma})}dc
\end{eqnarray*}
in which the last integration can be computed numerically and
$$h^o(c_i,k_i|\gamma,\phi, t_{io})\propto k_B(t_i|c_i,k_i)G_0(c_i,k_i)\propto [c_i|\gamma,\phi,t_{io}][k_i|c_i,\gamma,\phi,t_{io}]$$
where
$$[c_i|\gamma,\phi,t_{io}]\propto c_i t_i^{c_i-1}I_{(0,\phi)}(c_i)~~~~~i=1,...,n$$
and
$$[k_i|c_i,\gamma,\phi,t_{io}]\propto Gamma(.|2,\frac{1}{[\frac{1}{\gamma}+ln(1+t_i^{c_i})]}).$$
To sample from $h^o(c_i,k_i|\phi,\gamma,t_{io})$, we first draw a value from $[c_i|\gamma,\phi,t_{io}]$ using
slice sampling \cite{walker7}. Then draw a value for $k_i$ from the Gamma distribution with parameters 2 and $\frac{1}{[\frac{1}{\gamma}+ln(1+t_{io}^{c_i})]}$.

For rightly censored data $(t_{ic})$ the conditional posterior density of $(c_i,k_i)$ is
$$f(c_i,k_i|{(c_i,k_i);i\neq i'},\nu, \gamma, \phi,t_{ic})=\frac{q^c_0 h^c(c_i,k_i|\phi,\gamma,t_{ic})+\sum_{j=1}^{n^{*(i)}}n^{*(i)}_j q^c_j\delta_{c^*_j,k^*_j}}{q_0^c+\sum_{j=1}^{n^{*(i)}}n^{*(i)}_j q^c_j}$$
where $q_j^c=1-K_B(t_{ic}|c_j^*,k_j^*)$
\begin{eqnarray*}
q_0^c &=& \nu \int_0^\phi \int_0^\infty (1-K(t_{ic}|c,k))G_0(c,k)dcdk\\
&=& \frac{\nu}{\phi \gamma} \int_0^\phi \int_0^\infty\frac{e^{-\frac{k}{\gamma}}}{(1+t_{ic}^c)^k}dkdc\\
&=& \frac{\nu}{\phi \gamma} \int_0^\phi (\frac{1}{\gamma}+ln(1+t_{ic}^{c}))dc
\end{eqnarray*}
which can again be computed using numerical integration and
\begin{eqnarray*}
h^c(c_i,k_i|\gamma,\phi, t_{ic})&\propto& (1-K_B(t_i|c_i,k_i))G_0(c_i,k_i)\\
&\propto & [c_i|\gamma,\phi,t_{ic}][k_i|c_i,\gamma,\phi,t_{ic}]\\
&= & \frac{I_{(0,\phi)}(c_i)}{\phi \gamma} \frac{1}{\frac{1}{\gamma}+ln(1+t_{ic}^{c_i})} k_i e^{-k_i(\frac{1}{\frac{1}{\gamma}+ln(1+t_{ic}^{c_i})})}\\
&=&\frac{I_{(0,\phi)}(c_i)}{\phi \gamma} \frac{1}{\frac{1}{\gamma}+ln(1+t_{ic}^{c_i})}\times Gamma(k_i|2,\frac{1}{\frac{1}{\gamma}+ln(1+t_{ic}^{c_i})}).
\end{eqnarray*}
Using the slice sampling, we can draw a sample from $h^c(c_i,k_i|\phi,\gamma,t_{ic})$.

Now we can the update $(c_i,k_i)$ for $i=1,...,n$, for observed and censored data. Then in a general form, $(c^*_j,k_j^*)'s$ can be updated on $\phi,\gamma$ and t as the following
\begin{eqnarray}
f(c^*_j,k_j^*| \phi, \gamma, t, n^*)\nonumber
&\propto& G_0(c^*_j,k_j^*|\gamma,\phi)\prod_{\{io:s_{io}=j\}}k_B(t_{io}|c^*_j,k_j^*)\prod_{\{ic:s_{ic}=j\}}(1-K_B(t_{io}|c^*_j,k_j^*))\nonumber\\
&\propto & [c_j^*|\gamma,\phi,t_{io}][k_j^*|c_j^*,\gamma,\phi,t_{ic}]\prod_{\{ic:s_{ic}=j\} }\frac{1}{(1+t_{ic}^{c_j^*})^{k_j^*}}\nonumber\\
&\propto & {c_j^*}^{n_j^o}I_{(0,\phi)}(c_j^*)\prod_{\{io:s_{io}=j\}}\frac{t_{io}^{c_j^*-1}}{1+t_{io}^{c_j^*}}\times Gamma(n_j^o+1,B^*)
\end{eqnarray}
where $B^*=\sum_{\{io:s_{io}=j\}}(\frac{1}{\gamma}+ln(1+t_{io}^{c_j^*}))+\sum_{\{ic:s_{ic}=j\}}ln(1+t_{ic}^{c_j^*})$ and $n_j^o$ is the number of observed data which located in cluster j.
To generate a sample from equation (9), we just need to draw from the first part. Sampling from the gamma distribution is simple.
To sample from
$$[c_j^*|\phi,\gamma,t]\propto {c_j^*}^{n_j^o}I_{(o,\phi)}c_j^*\prod_{\{io:s_{io}=j\}}\frac{t_{io}^{c_j^*-1}}{1+t_{io}^{c_j^*}}$$
$$\propto {c_j^*}^{n_j^o}I_{(o,\phi)}c_j^*\prod_{\{io:s_{io}=j\}}(t_{io}^{c_j^*-1})\frac{1}{1+t_{io}^{c_j^*}}$$
we introduce auxiliary variables $W=\bigl\{w_{io};\{io:s_{io}=j\}\bigr\}$ such that
$$[c_j^*,W|\phi,t_{io}]={c_j^*}^{n_j^o}1_{(c_j^*\leq \phi)}\prod_{\{io:s_{io}=j\}} 1_{(w_{io}\leq \frac{t_{io}^{c_j^*-1}}{1+t_{io}^{c_j^*}})}.$$
By marginalization over the auxiliary variables, we get the $[c_j^*|\phi,t_{io}]$ for $j=1,...,n^*$.
Moreover, $w_{io}$'s are uniform on $(0,\frac{t_{io}^{c_j^*-1}}{1+t_{io}^{c_j^*}})$.
Therefore we have
$$[c_j^*|\phi,t]={c_j^*}^{n_j^o}I_{(B,\phi)}(c_j^*)$$
where $B=max\{0,\frac{ln(w_{io})}{1+t_{io}}\}$ and now drawing from $[c_j^*|\phi,t]$ is straightforward.

Afterwards, using the methods in \cite{escobar95}, we update $\phi,\gamma$ and $\nu$.
If we take u as a latent variable such that
$$[u|\nu,t]= Beta(\nu+1,n)$$
then
$$[\nu|u,n^*,t]=pGamma(a_\nu+n^*,b_\nu-log(u))+(1-p)Gamma(a_\nu+n^*-1,b_\nu-log(u))$$
where $p=\frac{a_\nu+n^*-1}{n(b_\nu-log(u))+a_\nu+n^*-1}$.

And finally to update $\phi$ we have
$$[\phi|c^*,k^*]=[\phi][c^*,k^*|\phi]=\frac{2b_\phi^2}{\phi^3}I_{(b_\phi,\infty)}(\phi)\prod_{j=1}^{n^*}\frac{1}{\phi}I_{(0,\phi)}(c^*)
=\frac{2b_\phi^2}{\phi^{n^*+3}}I_{(b^*,\infty)}(\phi)$$
where $b^*=max\{b_\phi,max_{1\leq j\leq n^*}c_j^*\}$.
So
$$[\phi|c^*,k^*]= Pareto(\phi|2+n^*,b^*)$$
and by repeating this technique , we can update $\gamma$
$$[\gamma|c^*,k^*]=[\gamma]\prod_{j=1}^{n^*}[k_j^*|\gamma]= IGamma(n^*+2,b_\gamma+\sum_{j=1}^{n^*}k_j^*).$$
Now, we can calculate all conditional distributions on the equation (3).

\section{Data illustration and case study}
We are going to work with two data sets to show the validity of the proposed model. First data set is a simulated data that we generate it from two mixture of Burr(XII) distributions. Another data set is a real data of breast cancer that was used in \cite{okasha}.
For each of the data sets we fit DPMM with different kernel distributions. Exactly Weibull, Log-normal and Burr(XII) distributions and then compare these models based on some indexes and also by their plots.

To see how well the models capture the true distribution, we use some goodness-of-fit(gof) metrics \cite{cheng13}.
Three of these metrics included,
$$ \frac{1}{n}\sum_{i=1}^{n}|\frac{Q_T(t_i)-\tilde{Q}_T(t_i)}{Q_T(t_i)}|,~\frac{1}{n}\sum_{i=1}^{n}|Q_T(t_i)-\tilde{Q}_T(t_i)|~and ~\frac{1}{n}\sum_{i=1}^{n}(Q_T(t_i)-\tilde{Q}_T(t_i))^2,$$
where $Q_T(t)$ can be any function of data.
We can calculate these metrics for the probability density, cumulative probability and hazard rate (HR) functions by replacing them with Q in the proposed gof functions.
To show the adaptivity of the model in survival applications, we apply it for comparison of two treatments.

\subsection{Simulated data}
Based on a finite mixture of Burr(XII) distributions, we simulate a complete life time data of the size n=200.
The finite mixture model is the mixture of two Burr(XII) distributions, exactly
$$pBurr(5,1)+(1-p)Burr(2,6)$$
where Burr(c,k) denotes the Burr(XII) distribution with the scale parameter c and the shape parameter k. We use p=0.2 to weight the Burr(XII) distributions.

As figure 1 shows simulated data has a bimodal probability density function that it is common in survival data. Also plots (c) and (d) of figure 1 illustrate cumulative and hazard rate functions of simulated data respectively.
\begin{figure}[t]
\begin{center}
\includegraphics[width=10cm,height=10cm]{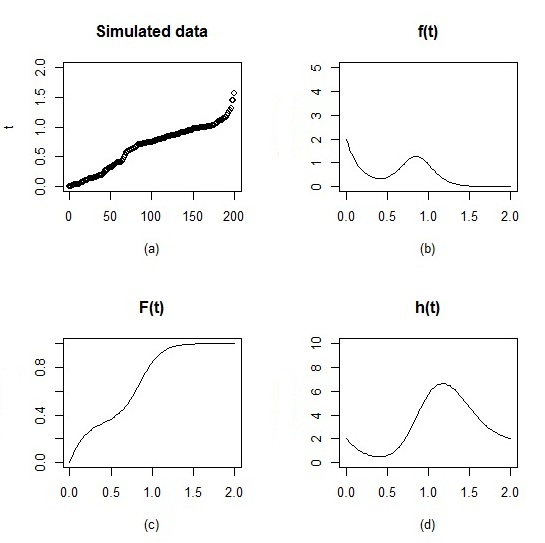}\\
\vspace{-0.2cm}
\caption{\footnotesize{Simulated data from mixture of Burr(XII) distributions (a)Empirical plotting position;(b)density function;(c) distribution function; (d) hazard function}}
\end{center}
\end{figure}
Table 1 illustrates indexes that we introduced in the beginning of this section and also we calculated values for DPMMs with different kernels. In this table
calculated indexes are for probability density functions of the models.
\begin{table}
\begin{center}
\begin{tabular}{|c|c|c|c|}
\hline
indexes & Weibull & Log-Noraml &  Burr(XII) \\
\hline
$\frac{1}{n}\sum_{i=1}^{n}|\frac{f_T(t_i)-\tilde{f_T}(t_i)}{f_T(t_i)}|$ & 1.7544 & 0.6346 & 0.3041\\
\hline
$\frac{1}{n}\sum_{i=1}^{n}|f_T(t_i)-\tilde{f_T}(t_i)|$  &157.1424 & 110.4019 & 26.8968\\
\hline
$\frac{1}{n}\sum_{i=1}^{n}(f_T(t_i)-\tilde{f_T}(t_i))^2$ & 0.7898 & 0.4587 & 0.0296 \\
\hline
\end{tabular}
\end{center}
\vspace{-0.4cm}
\caption{\footnotesize{Computed metrics for DPMMs with different kernels}}
\end{table}

The values in the table 1 signify that DPBMM has a better fitness for the simulated data,
also cumulative probability and hazard rate functions, they both proved preference of the DPBMM for this simulated data set.

Figure 2 shows the posterior mean of $f_T$, $F_T$ and $h_T$ respectively, using DPBM, DPWM and DPLNM models with the $\upsilon\sim Gamma(1,0.001)$. By \cite{kottas6} we know that with different choices of $\upsilon$'s parameters, there is a learning on $n^*$, so we select a approximately non-informative prior for $\upsilon$.In this figure density function of DPBMM fits the exact curve of simulated data very well.Also the cumulative distribution and hazard functions of DPBMM has a good fitness than other DPMMs.
\begin{figure}[t]
\begin{center}
\includegraphics[width=9.5cm,height=11cm]{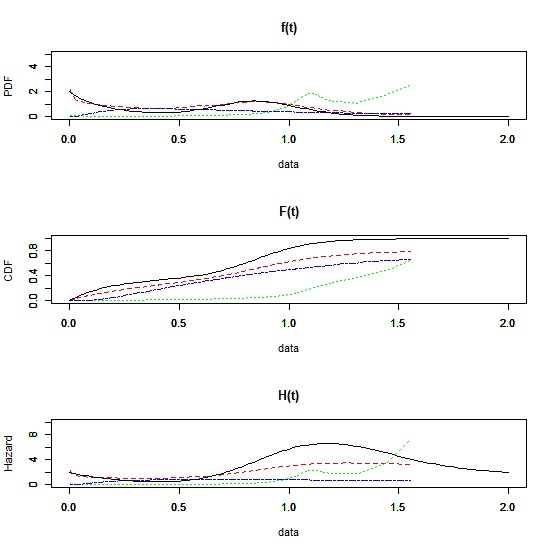}\\
\vspace{-0.2cm}
\caption{\footnotesize{(a) the density, (b) the cumulative distribution function and (c) the hazard function of simulated data with different kernels, solid line is true density, dashed line DPBMM, dotted line, DPWMM and dash-dotted line DPLNMM.}}\label{}
\end{center}
\end{figure}
\subsection{Case study}
To show the preference of DPBM model rather than DPLNM and DPWM models we apply these models for a real survival time data and compare them. We use a breast cancer data set that the data is obtained from \cite{okasha}. The data is the incidence and death dates of about 1000 breast cancer patients in Gaza strip within a period of 5 years starting from the being of 2009 to the end of 2013. The survival times for those patients that remained valid is 242 patients.

In figure 3 we plot the histogram of data in scale of 1000 days and the probability density of DPBM, DPLNM and DPWM models. The dashed line is DPBMM, the dotted line is DPLNMM and another one is DPWMM. The DPBMM captures the shape of the histogram very well
\begin{figure}[t]
\begin{center}
\includegraphics[width=9cm,height=10cm]{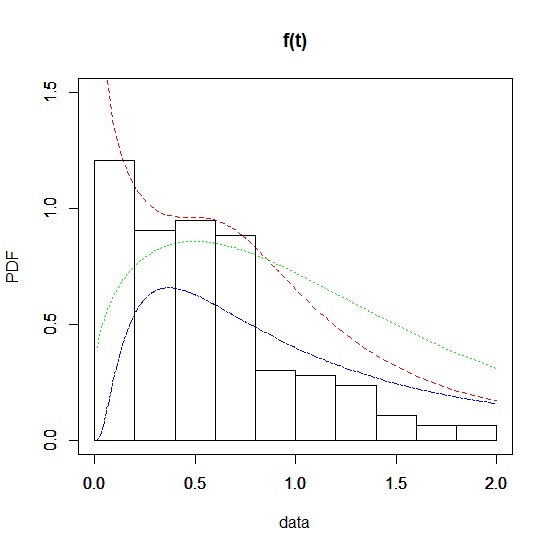}\\
\vspace{-0.2cm}
\caption{\footnotesize{Histogram of Ghaza data set and the estimated density function with DPBMM (dashed line), DPLNMM(dotted line) and DPLNMM(dash-dotted line) }}\label{}
\end{center}
\end{figure}
Figure 4 shows the reliability of data with step line and DPMMs with different kernels.
\begin{figure}[t]
\begin{center}
\includegraphics[width=9cm,height=10cm]{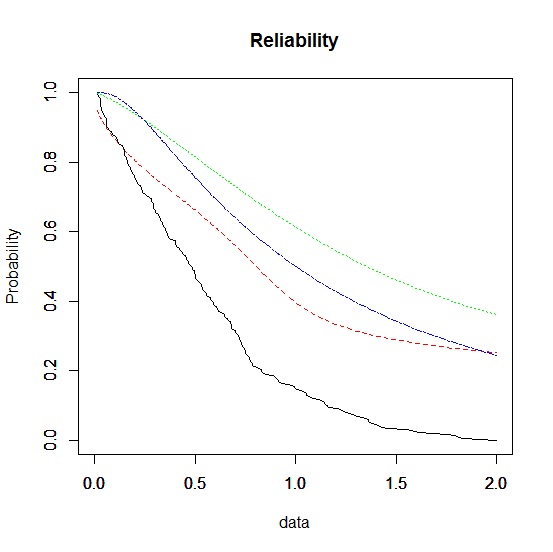}\\
\vspace{-0.2cm}
\caption{\footnotesize{Empirical reliability curve of Ghaza data set and the estimated reliability function with DPBMM (dashed line), DPLNMM(dotted line) and DPLNMM(dash-dotted line)}}\label{}
\end{center}
\end{figure}
As we see, DPBM model is the closest line to the reliability of data.
\begin{figure}[t]
\begin{center}
\includegraphics[width=9cm,height=10cm]{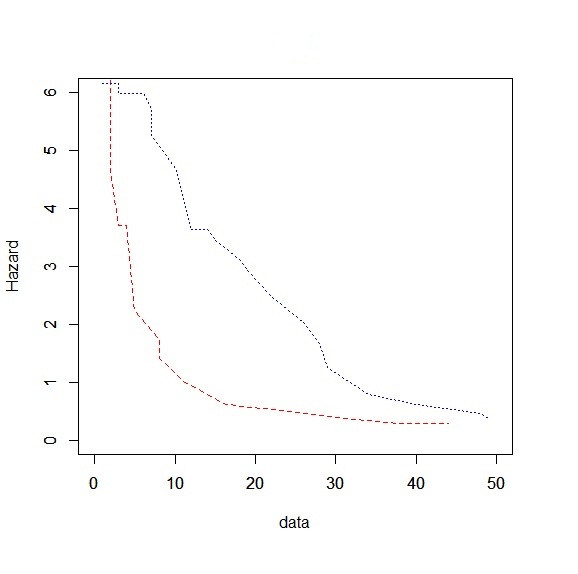}\\
\vspace{-0.2cm}
\caption{\footnotesize{Hazard rates for treatment A (dashed line) and treatment B (dotted line) with the mean of posterior in DPBMM}}
\end{center}
\end{figure}
\subsection{Comparison of two treatments}
Here we consider the leukemia patients data on remission times, in weeks that taken from \cite{Lawless}. Our purpose of
choosing this data is the comparison of survival functions across their populations with DPBMM. The data set involve two treatments, A and B, each of them included 20 patients. The data set is presented in table 2.
\begin{table}
\begin{center}
\begin{tabular}{|c c|c c|}
  \hline
  Treatment A & & Treatment B &\\
  \hline
  1& 3 & 1 & 1\\
  3&6&2&2\\
  7&7&3&4\\
  10 & 12&5& 8\\
  14 & 15 & 8 & 9\\
  18 & 19& 11 & 12\\
  22 & 26 & 14 & 16\\
  28+ & 29 & 18 & 21\\
  34 & 40 & 27+ & 31\\
  48+ & 49+ & 38+ & 44\\
  \hline
\end{tabular}
\end{center}
\vspace{-0.4cm}
\caption{\footnotesize{Leukemia patients data for two treatments A and B}}
\end{table}
\newline Lawless \cite{Lawless} tested the equality of the hazard functions of two treatments A and B based on classical test procedures and concluded that " there is no evidence of a difference in distributions".
In the another work Damien and walker \cite{Damien} compared these two treatments with a Bayesian nonparametric approach.
Their  approach did not assume any functional relationship between the distribution functions associated with the treatments and
yielded a result they regarded " far from conclusive of no
difference". But Kottas \cite{kottas6} employ  the DPWMM for each of  the underlying population distributions forcing no particular
relation between the corresponding distribution distribution functions.
He plots the 95$\%$ confidence interval for the mean of posterior and compare the difference between the treatments by posterior
means of density functions and hazard functions.

Here we apply the DPWMM to the two data set and calculated the mean of posterior for density and hazard functions.
Figure (5) shows the calculated quantities for the two treatment. As we see the difference between the treatments is obviously.

\section{Conclusion}
Here in this article we developed a mixture model for survival inference based on Burr XII as the kernel. Since we put priors on the space of parameter and the number of components is infinite, so we have a Bayesian nonparametric model. In this article we have applied our model i.e. Dirichlet process Burr(XII) mixture model for modeling of survival populations.
At first we applied the proposed model to the simulated data and then used it for modeling a real survival data. For both of them the DPBMM was fitted much more better than other models. Also for the data of two treatments the DPBMM showed the difference between two treatments.
For a further work we can propose using another kernel and introducing a new model.
%Also we can put another priors on the parameter space, especially the prior on the weights can be considered much more general than Dirichlet process, for example Polya tree or tail free process.

 \appendix{}
 \section{Choosing $b_\gamma$ and $b_\phi$ in the equation (8)}
 Since $\gamma \sim IG(2,b_\gamma)$ and $\phi \sim Pareto$ and by (7) we have
 $$c|\phi\sim Unif(0,\phi)$$
 $$k|\gamma\sim Exp(\gamma)$$
 so we can calculate the marginal distributions of c and k as bellow
 \begin{eqnarray}
[c]=\frac{2b_\phi^2}{3}\frac{2}{(max\{c,b_\phi\})^3}~~~~~~~~~~~~~~[k]=\frac{2b_\gamma^2}{(k+b_\gamma)^3}.
\end{eqnarray}
 Then the medians of c and k is obtained as the following
 $$m_c=\frac{3}{4}b_\phi~~~~~~~~~~~m_k=(\sqrt{2}-1)b_\gamma.$$

If we follow the methods that is given in \cite{kottas6}, we use the median and interquartile range of Burr(XII) distribution for obtaining a prior guess for (c,k) which denoted by $\tilde{c}$ and $\tilde{k}$.
\newline At first we know
$$F^{-1}_{B}(p)=((1-p)^{-1/k}-1)^{1/c}$$
and we have the quartiles by replacing $p=\frac{1}{4},\frac{2}{4},\frac{3}{4}$.
$$Q_1 = ((4/3)^{1/k}-1)^{1/c}~~~Q_2 = ((2)^{1/k}-1)^{1/c}~~~Q_3 = ((2)^{2/k}-1)^{1/c}$$
By putting equal the empirical and theoretical quartiles, the values of $\tilde{c}$ and $\tilde{k}$ are obtained and then to specify $b_\gamma$ and $b_\phi$, set medians of [c] and [k] equal to $\tilde{c}$ and $\tilde{k}$ respectively.





\end{document}